# Degree and connectivity of the Internet's scale-free topology*

Zhang Lian-Ming(张连明)[a)†], Deng Xiao-Heng(邓晓衡)[b)], Yu Jian-Ping(余建平)[c)], and Wu Xiang-Sheng(伍祥生)[a)]

[a)]*College of Physics and Information Science, Hunan Normal University, Changsha, 410081, China*

[b)]*Institute of Information Science and Engineering, Central South University, Changsha, 410083, China*

[c)]*College of Mathematics and Computer Science, Hunan Normal University, Changsha, 410081, China*

This paper theoretically and empirically studies the degree and connectivity of the Internet's scale-free topology at an autonomous system (AS) level. The basic features of scale-free networks influence the normalization constant of degree distribution $p(k)$. It develops a new mathematic model for describing the power-law relationships of Internet topology. From this model we theoretically obtain formulas to calculate the average degree, the ratios of the $k_{\min}$-degree (minimum degree) nodes and the $k_{\max}$-degree (maximum degree) nodes, and the fraction of the degrees (or links) in the hands of the richer (top best-connected) nodes. It finds that the average degree is larger for a smaller power-law exponent $\lambda$ and a larger minimum or maximum degree. The ratio of the $k_{\min}$-degree nodes is larger for larger $\lambda$ and smaller $k_{\min}$ or $k_{\max}$. The ratio of the $k_{\max}$-degree ones is larger for smaller $\lambda$ and $k_{\max}$ or larger $k_{\min}$. The richer nodes hold most of the total degrees of Internet AS-level topology. In addition, it is revealed that the increase rate of the average degree or the ratio of the $k_{\min}$-degree nodes has power-law decay with the increase of $k_{\min}$. The ratio of the $k_{\max}$-degree nodes has a power-law decay with the increase of $k_{\max}$, and

* Project supported by the National Natural Science Foundation of China (Grant No. 60973129, 60903058 and 60903168), the Specialized Research Fund for the Doctoral Program of Higher Education (Grant No. 200805331109), the China Postdoctoral Science Foundation (Grant No. 200902324) and the Program for Excellent Talents in Hunan Normal University, China (Grant No. ET10902)

† Corresponding author. E-mail: lianmingzhang@gmail.com

the fraction of the degrees in the hands of the richer 27% nodes is about 73% (the '73/27 rule'). Finally, empirically calculations are made, based on the empirical data extracted from Border Gateway Protocol, of the average degree, ratio and fraction using our method and other methods, and find that this method is rigorous and effective for Internet AS-level topology.

**Keywords**: scale-free networks, power-law distribution, Internet topology, average degree

## 1. Introduction

Over the past decade, there have been continuously emerging new theories and applications of complex networks. It has in fact become an important branch of various disciplines research, such as physics, informatics, biology, sociology, and so on [1][2]. The power-law distribution, hierarchy structure, rich-club phenomenon, disassortativity, self-similarity, community structure and other statistical properties of complex networks continues to have been unveiled [3]-[8]. Some topology models are presented for better depicting these properties, including the small-world and scale-free models, the local-world evolving network models, the module and hierarchical network models [8]-[12]. The relations between network topologies and synchronization behaviors in small-world networks or scale-free networks are investigated. The improved methods of control and synchronization and the models of phase synchronism of complex networks are given [13]-[18]. The spread mechanism and dynamics of the epidemic for adopting corresponding immunization strategy and the occurring mechanism, prevention and control of the cascading failure for improving the safety and reliability of complex networks are discussed [19][20]. Furthermore, the search and routing strategies are important topics in complex networks, including breadth-first search, random walk, maximum degree and greedy algorithms, and so on [21]-[23].

The Internet is an open man-made complex giant system and there have some advantages of its topological data compared with other complex networks. For instance, it is relatively easy to measure and obtain. Therefore, the Internet internationally was a typical example of a complex networks to be studied. However, with the rapid development of the large-scale Internet, the relation between network topology and network function has become one of the most important and difficult challenges offered by the Internet. The pioneering work [3], revealing the power-law distribution of autonomous system (AS)-level and router-level Internet topology, has inspired a great number of studies on topology characteristics, models, generators and roles of the Internet. Most network topology metrics have a determinant role in modeling the procedure of Internet topology. For example, assortativity coefficient, rich-club connectivity, coreness, average path length, average neighbor connectivity, clustering and betweenness are exactly equal or direct related to topology characteristics (disassortativity, rich- club phenomenon and small-world effect) [5][6][24]-[26]. Many models, such as extended scale-free model (SF), generalized linear preference model (GLP), positive feedback preferential model (PFP), dynamic and preferential model (DP), incremental edge addition and super-linear preferential attachment model (IEASPA) and multi-local world model (MLW) [27]-[32], can explain the growth of topology and develop some topology generators of the Internet, including Inet and Boston university representative Internet topology generator (BRITE) [33][34].

Inter order to better understand the dynamical behaviors of the Internet, to accurately depict the topological structure of the Internet and to availably control the efficient operation of the Internet, there should be further research on the statistical characteristics of Internet topology. As one of the basic metrics of the network, the average degree, $<k>$, is known as the mean value of the degrees for all nodes in a network. It can be expressed by $<k>=2m/n$, where $m$ refers as the number of links, and $n$ as the number of nodes. The average degree characterizes the global properties of the network and affects most topology characteristics. The networks with the same average degree can have different topologies. The average degree reflects the average connected performance (average connectivity) of the network and the greater value of the average degree, the more the neighbor nodes there are and the better the average connected performance is. For random networks [35], if $<k><1$, they compose of isolated nodes; if $<k>>1$, a connected network with a cluster appears; if $<k>>ln(n)$, network fully connects. Otherwise the edges and interference increase the value of the average degree, and search, routing and navigation become more complicated.

The average neighbor connectivity is a ratio to the average degree [26]. The average path length of the scale-free networks with some fixed average degree is bound by a polynomial in $log(n)$. The average degree has an influence on roles of the network. The network is likely to be more robust for larger average degree. Robustness against cascading failures has a positive correlation with the average degree [36]. The threshold of the infection spreading rate of scale-free networks is the value of the average degree divided by the second moment of the connectivity distribution [37]. However, the relations between the average degree and the topological parameters of the network play an important role in grasping mechanisms for the average degree impacting in what way the network structure and the network functions. Recently, based on the threshold preferential attachment model, Santiago and others analysed the impact of the affinity parameters and the topological parameters on the average degree and the highest degree [38]. Newman made an estimated calculation on the average degree of a discrete power-law distribution instead of a continual power-law distribution [39].

In this paper, based on the theoretical and empirical analysis of the Internet's scale-free topology, the findings are as follows. (1)The topological parameters of the Internet influence the average degree $<k>$, the ratios of the $k_{min}$-degree and $k_{max}$-degree nodes, and the fraction of the degrees (or links) in the hands of the richer nodes. The parameters include the power-law exponent $\lambda$, the minimum degree $k_{min}$ and the maximum degree $k_{max}$. (2)For For autonomous system (AS)-level Internet topology ($\lambda=2.25\pm0.01$, $k_{min}=1$), the average degree falls between 4.0 and 5.2, and the ratio of the $k_{min}$-degree nodes is about 60.8%~68.8%, and then the ratio of the $k_{max}$-degree nodes (hubs) is low. (3)The value of $r_{kmin}$ and the increase rate of $<k>$ have both a power-law decay with the increase of $k_{min}$, and the value of $r_{kmax}$ has power-law decay with the increase of $k_{max}$. (4)The AS-level Internet topology has the '73/27 rule'. Combined with the results in Ref. [39]-[44], the main different contributions of our works in this paper are as follows. Firstly, we introduce the maximum degree $k_{max}$ of Internet topology, present a new model and the calculation formula of the average degree. Secondly, we reveal three potential power laws of the average degree, the ratios $r_{kmin}$ of the $k_{min}$-degree nodes, and the ratios $r_{kmax}$ of the $k_{max}$-degree ones. Thirdly, we find the '73/27 rule' of Internet topology. The aforementioned findings and contributions are used for modeling the scale-free Internet, building the topology generator of large-scale networks. More importantly, they may have potential applications to traffic congestion control, information navigation, resource search and packet routing in various complex networks [45]-[47].

The rest of the paper is organized as follows. In Section 2 we discuss the mathematic models of complex networks and present a new model suitable for modeling scale-free Internet topology. In Section 3 we obtain the results of the analysis of the joint influence of the power-law exponent $\lambda$, the minimum degree $k_{min}$ and the maximum degree $k_{max}$ on the average degree $<k>$. In Section 4 we obtain the results of the analysis of the joint influence of $\lambda$, $k_{min}$ and $k_{max}$ on the ratios $r_{kmin}$ and $r_{kmax}$ of the $k_{min}$-degree nodes and the $k_{max}$-degree ones. In Section 5 we obtain the results the analysis of the joint influence of $\lambda$, $k_{min}$ and $k_{max}$ on the fraction of the degrees in the hands of the richest of the nodes. In Section 6 we demonstrate the results in the empirical data. Finally, Section 7 contains the summary of results and some concluding remarks.

## 2. A mathematic model of Internet's scale-free topology

In this section, we analyse the existing mathematic models of scale-free networks and develop a new model for the Internet's scale-free topology. For scale-free networks, the probability distribution $p(x)$ must satisfy the scale-free [39], namely,

$$p(ax)=bp(x). \quad (1)$$

For any $a$, $b$ is a constant. Assuming $p(1)p'(1)\neq 0$, and we infer the distribution $p(x)=p(1)x^{-\lambda}$, where $\lambda=-p(1)/p'(1)$, $p'(1)$ is the derivative of $p$ with $x=1$. Thus the continual power-law distribution is the only function satisfying the scale-free condition.

In real networks, the quantity $x$ refers to the node degree (discrete quantity). The network with the power-law degree distribution is also sometimes called the scale-free network. In other words, we have

$$p(k)=Ck^{-\lambda}, \quad (2)$$

where $k$ is the node degree, $p(k)$ is the probability of the $k$-degree nodes, and $\lambda$ is the power-law exponent (for the Barabási Albert-László (BA) scale-free network, $\lambda\approx 3$; for many real finite scale-free networks [40], $2\leq\lambda\leq 3$.), and $C$ is a normalization constant. The constant $C$ value, estimated from a continual power law distribution [40], can approximatively express by $C\approx(\lambda-1)k_{min}^{\lambda-1}$, where $k_{min}$ is the minimum degree satisfying the power-law distribution (for Internet [40], $k_{min}=1$.). This estimated approach has applied in many applications, owing to simple expressions for continual power-law distribution, but it gives sometimes poor results and should avoid [39][41][42]. It is fortunate the pure and discrete power-law distribution with the minimum degree can express by $p(k)=\xi^{-1}(\lambda,k_{min})k^{-\lambda}$, where $\xi(\lambda,k')= \sum_{k=k'}^{\infty}k^{-\lambda} = \sum_{k=0}^{\infty}(k+k')^{-\lambda}$ is the generalized or incomplete $\zeta$-function, $\zeta(\lambda)$ is the Riemann zeta function defined as $\sum_{k=1}^{\infty} k^{-\lambda}$, others define above.

For real finite Internet, the power-law distribution in the empirical data shows there is a minimum degree $k_{min}$ as well as a maximum degree $k_{max}$ ($>k_{min}$) of those nodes satisfying power-law distribution [41][43]. As a result, for values $k_{min} \leq k \leq k_{max}$, the constant $C$ in equation (2) gives by the normalization requirement that

$$1 = \sum_{k=k_{min}}^{k_{max}} p(k) = C \sum_{k=k_{min}}^{k_{max}} k^{-\lambda} = C(\sum_{k=k_{min}}^{\infty} k^{-\lambda} - \sum_{k=k_{max}}^{\infty} k^{-\lambda}) = C(\zeta(\lambda, k_{min}) - \zeta(\lambda, k_{max})).$$

Therefore, the degree distribution of Internet's scale-free topology can express by

$$p(k) = \frac{k^{-\lambda}}{\zeta(\lambda, k_{min}) - \zeta(\lambda, k_{max})}, \quad k = k_{min}, k_{min}+1, \ldots, k_{max}, \tag{3}$$

where $k_{min}$ and $k_{max}$, satisfying the power-law distribution, are not the smallest and largest values of the degree $k$ observed. The $p(k)$ is the probability of the $k$-degree nodes according with the power-law distribution, excluding those with values below $k_{min}$ and above $k_{max}$. The degree distribution expressed by equation (3) considers the impact of the maximum degree on it comparing with those in Ref. [39]-[41].

## 3. The average degree of Internet's scale-free topology

In the following, we theoretically study and numerically analyse the average degree of the Internet topology with the power-law behavior. For homogeneous networks, the node degrees of almost all the nodes are obviously about equal to the average degree of the network, and there is hardly any node being far beyond the average degree. For instance, almost all the node degrees of the Erdös-Rényi (ER) random graph with the Poisson distribution, being equal to the average degree, are $p(n-1)$, where $p$ is the connectivity probability. The node degrees of all nodes in the Watts and Strogatz (WS) small-world network are also close to the average degree [9]. The node degrees of most of the nodes are small in diverse networks, but only a few of the nodes have high degrees (hubs). The average degree of the BA scale-free network is $2m_0$, where $m_0$ is the number of the potential links in the BA scale-free model [10]. The number of the low-degree nodes in the BA networks is more than those of the high-degree ones.

For real finite Internet, there are the minimum degree $k_{min}$ and the maximum degree $k_{max}$, and so the calculation formula of the average degree $<k>$ of the Internet topology with the power-law behavior can also express by

$$<k> = \sum_{k=k_{min}}^{k_{max}} k\, p(k). \qquad (4)$$

Using equation (3) and equation (4), we can calculate the value of the average degree of the Internet as follows:

$$<k> = \sum_{k=k_{min}}^{k_{max}} \frac{k^{1-\lambda}}{\zeta(\lambda, k_{min}) - \zeta(\lambda, k_{max})} = \frac{\sum_{k=k_{min}}^{k_{max}} k^{1-\lambda}}{\zeta(\lambda, k_{min}) - \zeta(\lambda, k_{max})} = \frac{\zeta(\lambda-1, k_{min}) - \zeta(\lambda-1, k_{max})}{\zeta(\lambda, k_{min}) - \zeta(\lambda, k_{max})}. \qquad (5)$$

Thus the average degree $<k>$ is a function of the topology parameters: the exponent $\lambda$, the minimum degree $k_{min}$ and the maximum degree $k_{max}$ satisfying the power-law distribution.

Fig.1 shows the joint impact of the exponent, the minimum degree and the maximum degree on the average degree of Internet's scale-free topology. We see a straightforward dependency: the average degree is larger for smaller power-law exponent $\lambda$ and larger $k_{min}$ value and $k_{max}$ value.

Fig.1a shows the joint impact of $\lambda$ on $<k>$ for various $k_{min}$ and $k_{max}$. The values of $<k>$, curving inwards, decay with the increase of $\lambda$ values regardless of $k_{min}$ values and $k_{max}$ values, nearly converging to a minimum value for the same $k_{min}$ values. The decay rates in the average degree by the unit exponent increase with the increase of $k_{max}$ values for the same $k_{min}$ values, and increase with the increase of $k_{min}$ values for the same $k_{max}$ values. In the limit $\lambda \to 3.0$, the minimum limit values of the average degrees for the networks with $k_{min}=1$ are 1.37. However, in the limit $\lambda \to 2.0$, the values of the average degrees for the networks with $k_{max}=600, 1500, 2600$ are 4.24, 4.80, 5.13, respectively.

Fig.1b shows the joint impact of $k_{min}$ on $<k>$ for variety of $\lambda$ and $k_{max}$ in the log-log scale. The values of $<k>$, curving outwards, increase with the increase of $k_{min}$ values regardless of $\lambda$ values and $k_{max}$ values, around converging to a maximum value $k_{max}-1$ for the same $k_{min}$ values. The increase rates have a fast first increase with the increase of $k_{min}$ values, and they are larger for smaller $\lambda$ values and for larger $k_{max}$ values. For instance, if $k_{max}=1500$, the maximum values of $<k>$ are 1499 regardless of $\lambda$ values. If $k_{min}=1$, the values of $<k>$ for the networks with $\lambda=2.00, 2.25, 2.50, 2.75,$ and $3.00$ are 4.80, 2.71, 1.91, 1.55 and 1.37, respectively.

Fig.1c shows the joint impact of $k_{max}$ on $<k>$ for various $\lambda$ and $k_{min}$. The values of $<k>$ increase with the increase of $k_{max}$ values, and increase steeply for the regime with low-value $k_{max}$ and in-creases slightly for the regime with high-value $k_{max}$ regardless of $\lambda$ values and $k_{min}$ values. The increase rates of $<k>$ are larger for smaller $\lambda$ values and larger $k_{min}$ values, and have power-law decays with the increase of $k_{max}$ values. If $k_{max}=600$, the value of $<k>$ for the networks with $\lambda=2.00$, 2.25, 2.50, 2.75 and 3.00 are 4.24, 2.59, 1.89, 1.55, and 1.37, respectively. If $k_{max}=1500$, the matching values of $<k>$ are 4.80, 2.71, 1.91, 1.55 and 1.37, respectively.

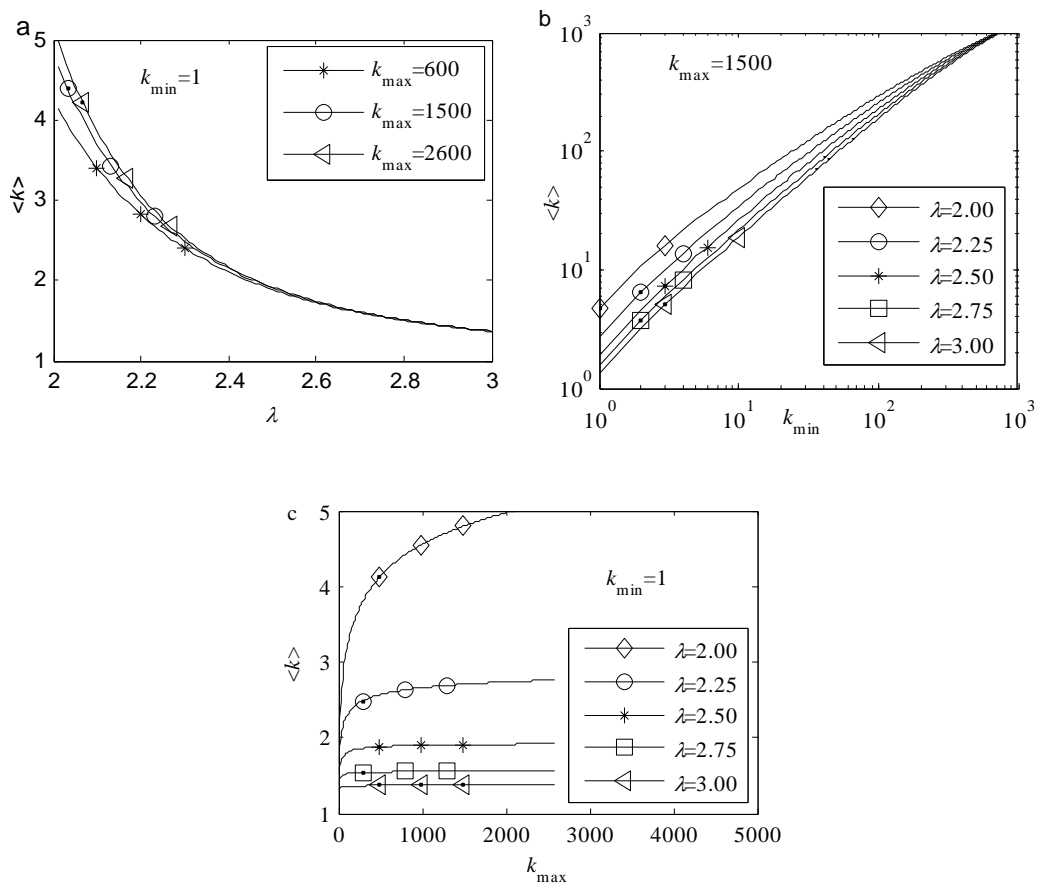

**Fig. 1.** The average degree of Internet's scale-free topology: (a) the average degree, $<k>$, as a function of the power-law exponent $\lambda$ for different values of the minimum degree $k_{min}$ and the maximum degree $k_{max}$, (b) $<k>$ as a function of $k_{min}$ for different $\lambda$ values and $k_{max}$ values in the log-log scale, (c) $<k>$ as a function of $k_{max}$ for different $\lambda$ values and $k_{min}$ values.

In summary, the topological parameters, such as the power-law exponent, the minimum degree and the maximum degree, influence the average degree of Internet's scale-free topology. The average degree is larger for smaller exponent, larger minimum and maximum degree, and the impacts of the smaller exponent and maximum degree and larger minimum degree are larger. Comparing with the near model: $<k> = (\lambda-1)/(\lambda-2)k_{min}$ (equation (11) in Ref. [39]), we find that our model (equation (5)) is more adaptable to a calculation of the average degree of the real finite Internet and other real finite complex networks.

## 4. The ratios of the $k_{min}$-degree nodes and the $k_{max}$-degree ones

In this section, we discuss the ratios of the $k_{min}$-degree nodes and the $k_{max}$-degree ones of Internet's scale-free topology. For BA scale-free networks, the number of the 1-degree nodes, the 2-degree ones and the 3-degree ones are 0, 0 and 40 percent of the number of all nodes in the network, respectively [10]. Based on the theoretical analysis, the results show that the number of the 1-degree nodes is more than or equal to 61 percent of the number of all nodes in a scale-free network [44]. In addition, the empirical analysis shows the number of the 1-degree nodes, 2-degree ones and 3-degree ones are 26, 38 and 14 percent of all nodes of the AS-level Internet topology, respectively [29].

In real finite scale-free Internet, let $n_k$, $p(k)$ and $N$ mark the number of the degree-$k$ nodes, the likelihood of the degree-$k$ nodes and the number of all nodes satisfying the power-law distribution, respectively, giving

$$n_k = Np(k), \qquad k=k_{min}, \ldots, k_{max}. \tag{6}$$

Supposing that $n_{k_{min}} \neq 0$, we have

$$n_k = n_{k_{min}} p(k)/p(k_{min}), \quad N = \sum_{k=k_{min}}^{k_{max}} n_k. \tag{7}$$

Using equation (3) and equation (7), and then we obtain

$$N/n_{k_{min}} = k_{min}^{\lambda} \sum_{k=k_{min}}^{k_{max}} k^{-\lambda} = k_{min}^{\lambda}(\zeta(\lambda,k_{min})-\zeta(\lambda,k_{max})). \tag{8}$$

Equation (8) yields

$$n_{k_{min}} = \frac{N}{k_{min}^{\lambda}(\zeta(\lambda,k_{min})-\zeta(\lambda,k_{max}))}. \tag{9}$$

Let $r_{kmin} = n_{k_{min}}/N$, we hold the ratio $r_{kmin}$ of the $k_{min}$-degree nodes of Internet's scale-free topology as follows

$$r_{kmin} = k_{min}^{\lambda}(\zeta(\lambda,k_{min})-\zeta(\lambda,k_{max}))^{-1}. \tag{10}$$

In a similar way, we gain also the ratio $r_{kmax}$ of the $k_{max}$-degree nodes, namely

$$r_{kmax} = k_{max}^{\lambda}(\zeta(\lambda,k_{min})-\zeta(\lambda,k_{max}))^{-1}. \tag{11}$$

Thus the ratios of the $k_{min}$-degree nodes and the $k_{max}$-degree ones are both as the functions of the topological parameters: the power-law exponent $\lambda$, the minimum degree $k_{min}$ and the maximum degree $k_{max}$ of Internet's scale-free topology.

Fig.2 shows the joint impact of the power-law exponent $\lambda$, the minimum degree $k_{min}$ and the maximum degree $k_{max}$ on the ratio $r_{kmin}$ of the $k_{min}$-degree nodes. We see a straightforward dependency: the value of $r_{kmin}$ is larger for larger $\lambda$ value and smaller $k_{max}$ value, and the value of $r_{kmin}$ is larger for smaller $k_{min}$ value for the low-value $k_{min}$ and the value of $r_{kmin}$ is larger for larger $k_{min}$ value for the large-value $k_{min}$.

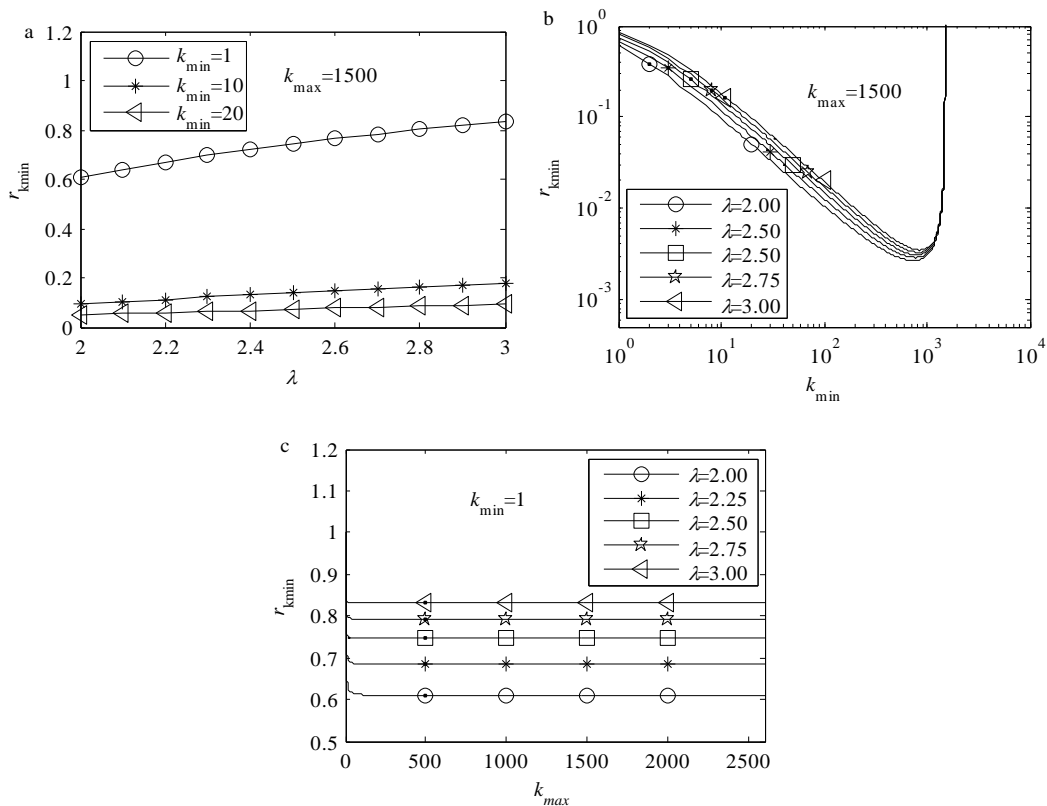

**Fig. 2.** The ratios of the $k_{min}$-degree nodes: (a) the ratio, $r_{kmin}$, as a function of the power-law exponent $\lambda$ for different values of the minimum degree $k_{min}$ and the maximum degree $k_{max}$, (b) $r_{kmin}$ as a function of $k_{min}$ for different $\lambda$ values and $k_{max}$ values in the log-log scale, (c) $r_{kmin}$ as a function of $k_{max}$ for different $\lambda$ values and $k_{min}$ values.

Fig.2a shows the joint impact of $\lambda$ on $r_{kmin}$ for different $k_{min}$ values and $k_{max}$ values. The ratios almost linearly increase with the increase of $\lambda$ values regardless of $k_{min}$ values and $k_{max}$ values. The increase rates of $r_{kmin}$ are smaller for larger $k_{min}$ values and smaller $k_{max}$ values. The values of $r_{kmin}$ for the networks with $\lambda$=2.00, 2.25, 2.50, 2.75 and 3.00, for instance, are equal to 61%, 68%, 75%, 79% and 83%, respectively, if $k_{min}$=1 and $k_{max}$=1500. In the limit $\lambda \to 3.0$, the ratios converge to-ward a maximum value for the same $k_{min}$ values. The maximum value of the ratios $r_{kmin}$ for the networks with $k_{min}$=1, 10 and 20 are 83%, 18% and 10%, respectively.

Fig.2b shows the joint impact of $k_{min}$ on $r_{kmin}$ for different $\lambda$ values and $k_{max}$ values in the log-log scale. The histograms are first straight lines with negative slopes regardless of $\lambda$ values and $k_{max}$ values for low-value $k_{min}$, meaning the ratios $r_{kmin}$ have power-law decays with the increase of $k_{min}$ values. In other words, the ratio of the $k_{min}$-degree nodes expresses by $r_{kmin} \propto k_{min}^{-\gamma}$. We find the exponents $\gamma$ are from -0.8 to -1 for the scale-free networks with $2 \leq \lambda \leq 3$, and $\gamma$ are larger for smaller $\lambda$ values and larger $k_{max}$ values. The ratios arrive at a minimum value if the $k_{min}$ values are equal to, or greater than half of $k_{max}$ values. For instance, if $\lambda$ values are 2.00, 2.25, 2.50, 2.75 and 3.00, the minimum values of $k_{min}$ for the networks with $k_{max}$=1500 is 0.0027, 0.0027, 0.0029, 0.0031, 0.0033 and 0.0035. The values of $k_{min}$ are 750, 784, 814, 841 and 866, respectively. However, the ratios increase with the increase of $k_{min}$ values for large-value $k_{min}$, and the increase rates of the ratios are larger for larger $k_{min}$ values.

Fig.2c shows the joint impact of $k_{max}$ on $r_{kmin}$ for different $\lambda$ values and $k_{min}$ values. The results show the influences of $k_{max}$ on $r_{kmin}$ display two different regimes regardless of $\lambda$ values and $k_{min}$ values. The ratios converge toward a minimum value with the decrease of $k_{max}$ values. The ratios have a sharp decrease with the increase of $k_{max}$ values in the first regime (smaller $k_{max}$ values). Then they come up to an inflection point, and hold all along in the second regime (larger $k_{max}$ values). The ratios in the inflection points are larger for larger $\lambda$ values and smaller $k_{min}$ values. For instance, the values of the ratios for the networks with $\lambda$=2.00, 2.25, 2.50, 2.75 and 3.00, for $k_{max}$=15, are 63.5%, 69.8%, 75.2%, 79.7% and 83.4%, respectively. For the networks with $k_{max}$=1500, they are 60.8%, 68.5%, 74.5%, 79.4% and 83.2%, respectively.

Fig.3 shows the joint impact of the power-law exponent $\lambda$, the minimum degree $k_{min}$ and the maximum degree $k_{max}$ on the ratio $r_{kmax}$ of the $k_{max}$-degree nodes. We see a straightforward dependency: the value of $r_{kmax}$ is larger for smaller $\lambda$ value and $k_{max}$ value and larger $k_{min}$ value.

Fig.3a shows the joint impact of $\lambda$ on $r_{kmax}$ for different $k_{min}$ values and $k_{max}$ values. The ratios have exponential decays with the increase of $\lambda$ values regardless of $k_{min}$ values and $k_{max}$ values. The decay rates of $r_{kmax}$ are larger for larger $k_{min}$ values and smaller $k_{max}$ values. The ratios $r_{kmax}$ for the networks with $\lambda$=2.00, 2.25, 2.50, 2.75 and 3.00, for instance, are equal to $0.27 \times 10^{-6}$, $0.49 \times 10^{-7}$, $0.86 \times 10^{-8}$, $0.15 \times 10^{-8}$ and $0.25 \times 10^{-9}$, respectively, if $k_{min}$=1 and $k_{max}$=1500.

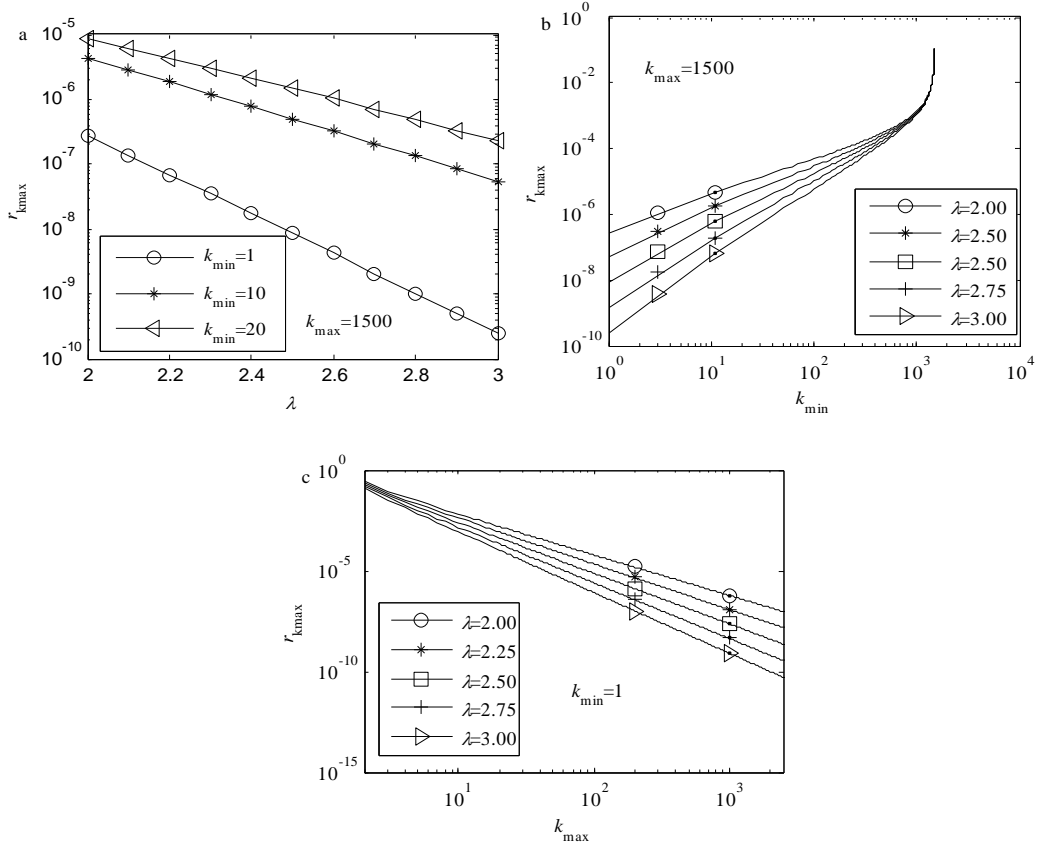

**Fig. 3.** The ratios of the $k_{max}$-degree nodes: (a) the ratios, $r_{kmax}$, as a function of the power-law exponent $\lambda$ for different values of the minimum degree $k_{min}$ and the maximum degree $k_{max}$ in the semi-log scale, (b) the ratio as a function of $k_{min}$ for different values of $\lambda$ and $k_{max}$ in the log-log scale, (c) the ratio as a function of $k_{max}$ for different values of $\lambda$ and $k_{min}$ in the log-log scale.

Fig.3b shows the joint impact of $k_{min}$ on $r_{kmax}$ for different $\lambda$ values and $k_{max}$ values in the log-log scale. The ratios have first slight increase with the increase of $k_{min}$ values regardless of $\lambda$ values and $k_{max}$ values, and the increase rates of $r_{kmax}$ become larger and larger if $k_{min}$ values near to $k_{max}$ values. The increase rates of the ratios for the networks with $k_{min}=20$ and 10 are 2.24 and 1.87 times as large as that of the ratio for the networks with $k_{min}=1$, respectively.

Fig.3c shows the joint impact of $k_{max}$ on $r_{kmax}$ for different $\lambda$ values and $k_{min}$ values in the log-log scale. The results show that histograms are steep curves for low-value $k_{max}$ and straight lines with negative slopes for large-value $k_{max}$ regardless of $\lambda$ values and $k_{min}$ values. This means the ratios have first sharp decays and then do power-law decays with the increases of $k_{max}$ values. The exponents of the power-law decays are equal to ones of the matching power-law degree distributions, if $k_{min}=1$. They reduce with the increase of $k_{min}$ values. For instance, the exponents of the power-law decays of the ratios for the networks with $\lambda=2.25$ and large-value $k_{max}$ are 2.25.

To sum up, the topological parameters influence the ratio $r_{kmin}$ of the $k_{min}$-degree nodes and the ratio $r_{kmax}$ of the $k_{max}$-degree ones of Internet's scale-free topology. The value of $r_{kmin}$ is larger for larger $\lambda$ value, smaller $k_{max}$ value and smaller $k_{min}$ value for the networks with low-value $k_{min}$, but the value of $r_{kmin}$ is larger for larger $k_{min}$ value for the networks with large-value $k_{min}$. The value of $r_{kmax}$ is larger for smaller $\lambda$ value and $k_{max}$ value, and larger $k_{min}$ value. For real finite Internet with $k_{min} \ll k_{max} \ll \infty$, we find the networks have two power laws: (1) $r_{kmin} \propto k_{min}^{-\gamma}$, $0.8 \leq \gamma \leq 1$; (2) $r_{kmax} \propto k_{max}^{-\lambda}$, $\lambda$ is the exponent of the power-law degree distribution.

## 5. The fraction of the degrees in the hands of the richest of the nodes

In this section, we study the fraction of the degrees (or links) in the hands of the richest of the nodes based on the model of Internet's scale-free topology. In real scale-free networks, the degrees of a few nodes are high, and the degrees of most of the nodes are low. For example, the top 20% of web sites obtain about 70% of all web sites, and the largest 10% of US cites house about 60% of the country's total population [39].

For real finite AS-level Internet topology [48], let us set $n_{>k}$ be the number of the nodes. The degree has a value greater than or equal to $k$. Then let us set $d_{>k}$ be the number of the total degrees in the hands of those nodes, giving

$$n_{>k} = \sum_{k}^{k_{max}} Np(k) = N\sum_{k}^{k_{max}} p(k), \qquad (12)$$

$$d_{>k} = \sum_{k}^{k_{max}} Nkp(k) = N\sum_{k}^{k_{max}} kp(k). \qquad (13)$$

Let $R_{nodes}$ be the ratio of $n_{>k}$ to the number $N$ of the nodes excluding those with values below $k_{min}$ and above $k_{max}$. Then let $R_{degrees}$ be the ratio of $d_{>k}$ to the number of total degrees in the hands of $N$ nodes, we have

$$R_{nodes} = \frac{n_{>k}}{N} = \sum_{k}^{k_{max}} p(k), \qquad (14)$$

$$R_{degrees} = \frac{d_{>k}}{N\sum_{k=k_{min}}^{k_{max}} kp(k)} = \frac{\sum_{k}^{k_{max}} kp(k)}{\sum_{k=k_{min}}^{k_{max}} kp(k)}. \qquad (15)$$

Using equation (3), equation (14) and equation (15), thus leading to

$$R_{nodes} = \frac{\zeta(\lambda, k) - \zeta(\lambda, k_{max})}{\zeta(\lambda, k_{min}) - \zeta(\lambda, k_{max})}, \qquad (16)$$

$$R_{\text{degrees}} = \frac{\zeta(\lambda-1,k) - \zeta(\lambda-1,k_{\max})}{\zeta(\lambda-1,k_{\min}) - \zeta(\lambda-1,k_{\max})} . \tag{17}$$

The equations (16) and (17) characterize the relation of the fraction $R_{\text{nodes}}$ of the richest nodes with the fraction $R_{\text{degrees}}$ of the total degrees in the hands of the richest ones.

Fig.4 shows the form of the curve of $R_{\text{degrees}}$ against $R_{\text{nodes}}$ for various values of the power-law exponent $\lambda$, the minimum degree $k_{\min}$ and the maximum degree $k_{\max}$. We see a straightforward dependency: the value of $R_{\text{degrees}}$ is larger for larger $R_{\text{nodes}}$ value regardless of the value of $\lambda$, $k_{\min}$ and $k_{\max}$, and $R_{\text{degrees}}$ has a first rapid increase and a tail-long slow increase with the increase of $R_{\text{nodes}}$ value.

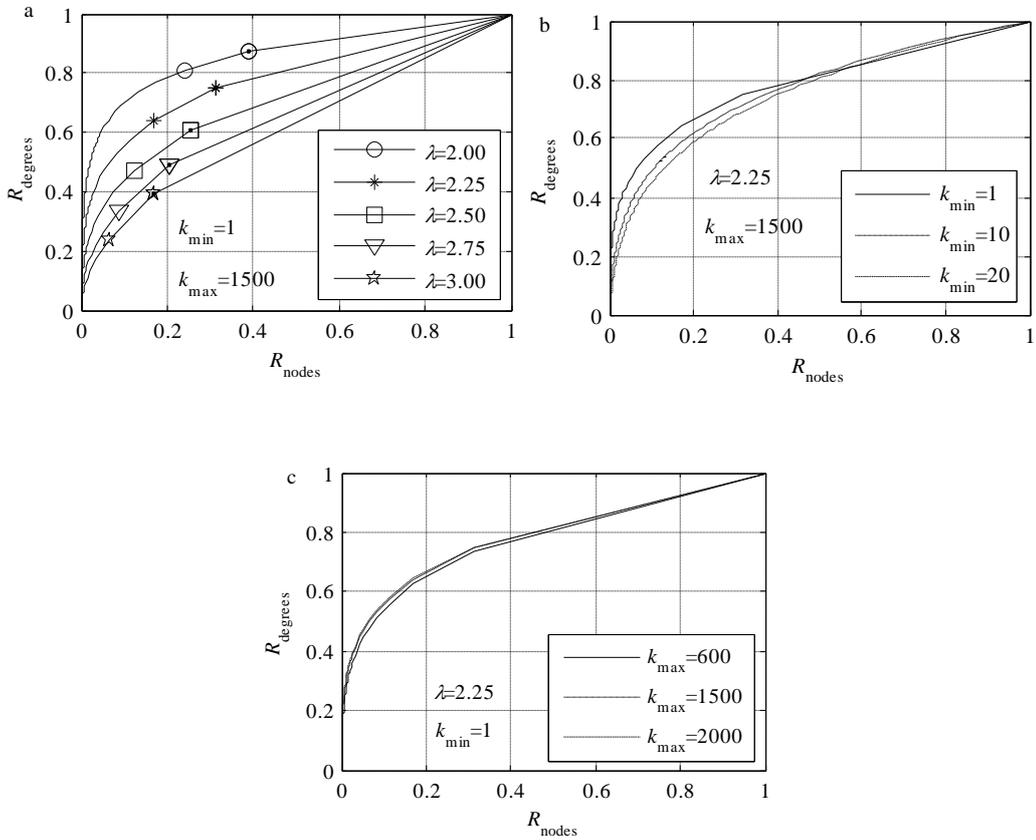

**Fig. 4.** The forms of the curves of $R_{\text{degrees}}$ as a function of $R_{\text{nodes}}$: (a) the curves of $R_{\text{degrees}}$ as a function of $R_{\text{nodes}}$ for the networks with $k_{\min}=1$, $k_{\max}=1500$ for different $\lambda$ values, (b) the curves of $R_{\text{degrees}}$ as a function of $R_{\text{nodes}}$ for the networks with $\lambda=2.25$, $k_{\max}=1500$ for different $k_{\min}$ values, (c) the curves of $R_{\text{degrees}}$ as a function of $R_{\text{nodes}}$ for the networks with $\lambda=2.25$, $k_{\min}=1$ for different $k_{\max}$ values.

Fig.4a shows the joint impact of the power-law exponent $\lambda$ on the fraction $R_{\text{degrees}}$ in the hands of the fraction $R_{\text{nodes}}$ for the networks with the same values of the minimum degree ($k_{\min}=1$) and the same values of the maximum degree ($k_{max}$=1500). The values of $R_{degrees}$ for the same values of $R_{nodes}$ are larger for smaller $\lambda$ values. As $R_{nodes}$ increases, $R_{degrees}$ increases rapidly at early stage, and then the increase rate of $R_{degrees}$ becomes smaller regardless of $\lambda$ values. The joint impacts of $k_{min}$ and $k_{max}$ on the values of the fraction $R_{degrees}$ against $R_{nodes}$ are similar to that of $\lambda$. For the same $R_{nodes}$ values, the values of $R_{degrees}$ are larger for smaller $k_{min}$ values (Fig.4b) and for larger $k_{max}$ values (Fig.4c). The curvatures of the curves are larger for smaller $\lambda$ values and $k_{min}$ values, and larger $k_{max}$ values. For instance, the top 17% richer nodes hold about 64% degrees (or links) if $\lambda$=2.25, $k_{min}$=1 and $k_{max}$=1500. If $2<\lambda<3$, the top 17% best-connected nodes hold about 40%~80% degrees for various values of $k_{min}$ and $k_{max}$. On the other hand, we find there is the '73/27 rule' if $\lambda$=2.25± 0.01, about 73% of the degrees are in the hands of the richest 27% of all nodes.

In summary, based on the theoretical analysis, we infer that most of the number of the degrees of the scale-free networks is in the hands of a few nodes of networks. The gaps between the rich (high-degree nodes) and the poor (low-degree ones) are larger for smaller exponents and minimum degrees, and larger maximum degrees. Interestingly, our results show the scale-free Internet has the '73/27 rule'.

## 6. The results and analysis in empirical data

### 6.1. The empirical data

BGP routing information is one of the important data sources of Internet topology at the AS level. Route Views project of University of Oregon collects BGP routing information from Route Views Collectors which place in several different locations all over the world around Internet [26]. In this paper, the data sources, being available from Ref. [49], apply to build the AS-level Internet topology. The data sources consist of the eight sets of data: AS1, AS2, AS3, AS4, AS5, AS6, AS7 and AS8.

After analyzing these data sources, we find there are some looping back links and some overlapping backward links. Firstly, we remove the looping and overlapping back links in the eight sets of data, and obtain an undirected and unweighted graph of Internet topology at the AS level. Secondly, based on maximum likelihood estimate [42][43], we hold the theoretical values of the basic metrics of the Internet topology with the power-law behavior: the power-law exponent $\lambda$, the min-imum degree $k_{min}$, the maximum degree $k_{max}$. The metrics are given in the Tab.1. The number of the nodes: $n$, the number of the links: $m$, the minimum degree: $k_{min0}$, the maximum degree $k_{max0}$ and the average degree: $<k_0>$ are some basic metrics of the graph after the first step.

**Tab.1.** The values of the topological metrics of Internet

|  | AS1 | AS2 | AS3 | AS4 | AS5 | AS6 | AS7 | AS8 |
|---|---|---|---|---|---|---|---|---|
| $n$ | 12694 | 7690 | 8689 | 8904 | 8063 | 10476 | 12694 | 12741 |
| $m$ | 26559 | 15413 | 17709 | 17653 | 16520 | 21113 | 26559 | 26888 |
| $k_{min0}$ | 1 | 1 | 1 | 1 | 1 | 1 | 1 | 1 |
| $k_{max0}$ | 2566 | 1713 | 1911 | 1921 | 1833 | 2274 | 2566 | 2557 |
| $<k_0>$ | 4.18 | 4.01 | 4.08 | 3.97 | 4.10 | 4.03 | 4.18 | 4.22 |
| $\lambda$ | 2.24 | 2.26 | 2.25 | 2.25 | 2.26 | 2.25 | 2.24 | 2.24 |
| $k_{min}$ | 1 | 1 | 1 | 1 | 1 | 1 | 1 | 1 |
| $k_{max}$ | 2031 | 1225 | 1378 | 1417 | 1289 | 1666 | 2031 | 2051 |

## 6.2. The results and analysis

In the following, we calculate and analyse the values of $<k>$, $r_{kmin}$, $r_{kmax}$ and the fraction of $R_{degrees}$ of the degrees in the hands of the richest of the fraction $R_{nodes}$ of the nodes in the empirical data as mentioned above.

Firstly, we calculate the theoretical values of the average degrees $<k>$ of the empirical data by our method (equation (5)) and Newman method (equation (11) in Ref. [39]), respectively, and compare the theoretical values and the real values $<k_0>$ (see Tab.1) of the average degree of the eight sets of the empirical data. The results from the equation (5) with $k_{max}$ in Tab.1 are better than those from Newman method. But the errors still come up to 18.8% ±9.0%. Through analysis and study, we find the errors from the equation (5) cause by the noncontinuity of the power-law degree distribution in the real Internet. We cancel the errors by network scale $n$ replacing the maximum degree $k_{max}$ in the equation (5). The theoretical values and the real values of the average degree of the eight sets of the empirical data and the errors of the different methods plot in Fig.5.

The results show the theoretical values of $<k>$, got by our method, are less than those of $<k>$ by Newman method. But they are greater than the real values of the average degree of the empirical data for AS1, AS2, …, AS8 (see Fig.5a). The values of the average degree got by three above-mentioned methods are 4.00±0.24, 5.02±0.18 and 4.10±0.13, respectively, this is, one node of the AS-level Internet topology has 4 or 5 1-hop neighbor nodes on average. The errors of our method, being 2.5%±5.8%, are less than 22.7%±4.4% of Newman method (see Fig.5b). The node degrees of the power-law distribution in the two theoretical methods are some successive integers (as defined in equation (3)), but those of the empirical data are nonconsecutive integers. In addition, Newman method includes the node degrees with $k>k_{max}$. The reasons may be to lead to most of the errors. We confirm the errors of our method, with the increase of the network size of the AS-level Internet topology, reduce less and less.

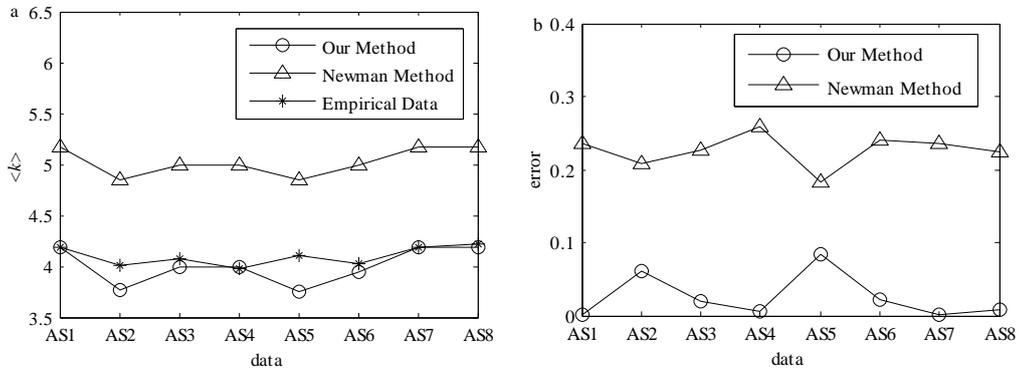

**Fig. 5.** The average degree and the errors of the different methods: (a) the theoretical and empirical values of the average degrees got by the different methods, (b) the values of the errors of the different methods.

Secondly, we calculate the theoretical values of the ratios $r_{kmin}$ of the $k_{min}$-degree nodes of the empirical data by our method (equation (10)) and XPP method (equation (15) in Ref. [44]). We compare the theoretical values and the real values of the ratios $r_{kmin}$ of the $k_{min}$-degree nodes of the empirical data. The values of the ratios $r_{kmin}$ of the $k_{min}$-degree nodes and the errors of these three methods plot in Fig.6.

In the empirical data, the values of the minimum degree of the eight empirical data are all equal to 1 (see Tab.1). In other words, $r_{kmin}$ are just the ratios of the leaf nodes (1-degree nodes) of the AS-level Internet topology. In addition, we analyse also the ratios of the low-degree nodes in the empirical data. The results show the ratios of the 1-degree, 2-degree and 3-degree nodes of the empirical data are 34.6%±2.4%, 40.6%±2.4% and 11%, respectively. We renormalize the ratio of the 1-degree, 2-degree and 3-degree nodes according to the power-law distribution. The results show the real values of the ratio of the 1-degree nodes for the eight sets of the empirical data are 67.1%, 65.8%, 65.7%, 67.2%, 65.8%, 67.2%, 67.1% and 67.1%, respectively.

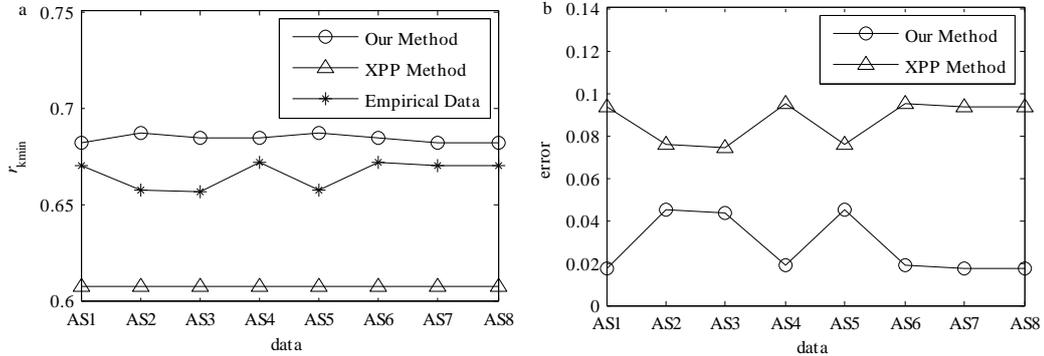

**Fig. 6.** The ratios $r_{kmin}$ of the $k_{min}$-degree nodes and the errors of the different methods: (a) the theoretical and empirical values of $r_{kmin}$ got by the different methods, (b) the values of the errors of the different methods.

Fig.6a shows the theoretical values, based on our method and XPP method, of the ratios of the $k_{min}$-degree (1-degree) nodes of the empirical data are 68.5%±0.3% and 60.8%. The renormalized value of the empirical data is 66.6%±1.0%. The joint impact of the exponent $\lambda$, the minimum degree $k_{min}$ and the maximum degree $k_{max}$ on the ratio of the $k_{min}$-degree nodes considers, so our method is more concrete. But our method is effective, and the error is 2.8%±1.8%, and it is less than 8.7%±1.3% of XPP method (see Fig.6b). Fig.7 shows the theoretical values of the ratios of the $k_{max}$-degree nodes of AS1, AS2, …, AS8 is $4.61 \pm 2.60 \times 10^{-8}$. The ratio is less than the real value of the ratio, being $1.02 \pm 0.28 \times 10^{-4}$, of the $k_{max}$-degree nodes of the empirical data.

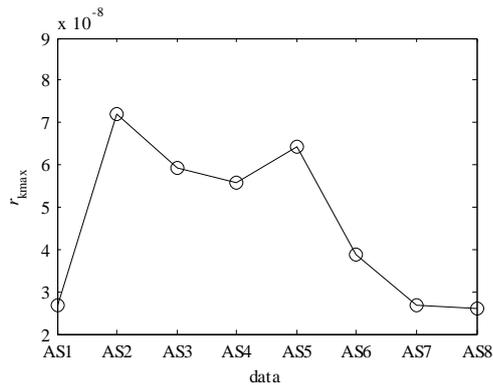

**Fig. 7.** The ratio $r_{kmax}$ of the $k_{max}$-degree nodes.

Finally, we calculate the theoretical values of the fraction $R_{degrees}$ of the degrees in the hands of top 20% and 27% of all nodes of the empirical data by our method (equation (16) and (17)) and Newman method (equation (29) in Ref. [39]). We compare the theoretical values and the real val-ues of the fraction of the degrees in the hands of those richest nodes of the empirical data. The theoretical values of $R_{degrees}$ in the hands of top 20% best-connected nodes by our method and Newman method are about 67.6% and 72.6% ±0.8%, respectively. The real values of the ratios of the degrees in the hands of the richest 20% nodes of the empirical data are about 68.8% ±0.6% (see Fig.8a). The errors of our method and Newman method are 1.7% ±0.9% and 5.6% ±1.3%, respectively (see Fig.8b). Newman method is perfect, and neglects the joint impact of the minimum degree and the maximum degree on the fraction $R_{degrees}$. Our method is suitable for analyzing the real finite Internet's scale-free topology. Further onwards, we reveal the '73/27 rule' of the AS-level Internet topology, this is, the top 27% best-connected nodes hold about 73% degrees of all nodes satisfying power-law behavior. The theoretical and empirical results show the values of $R_{degrees}$ in the hands of 27% nodes by our method, Newman method and the empirical data is 71.7% ±1.3%, 77.1% ±0.7% and 73.1% ±0.5%, respectively (see Fig.8c). The error of our method is 1.8% ±1.5%, which is fewer than 5.5% ±1.2% of Newman method (see Fig.8d).

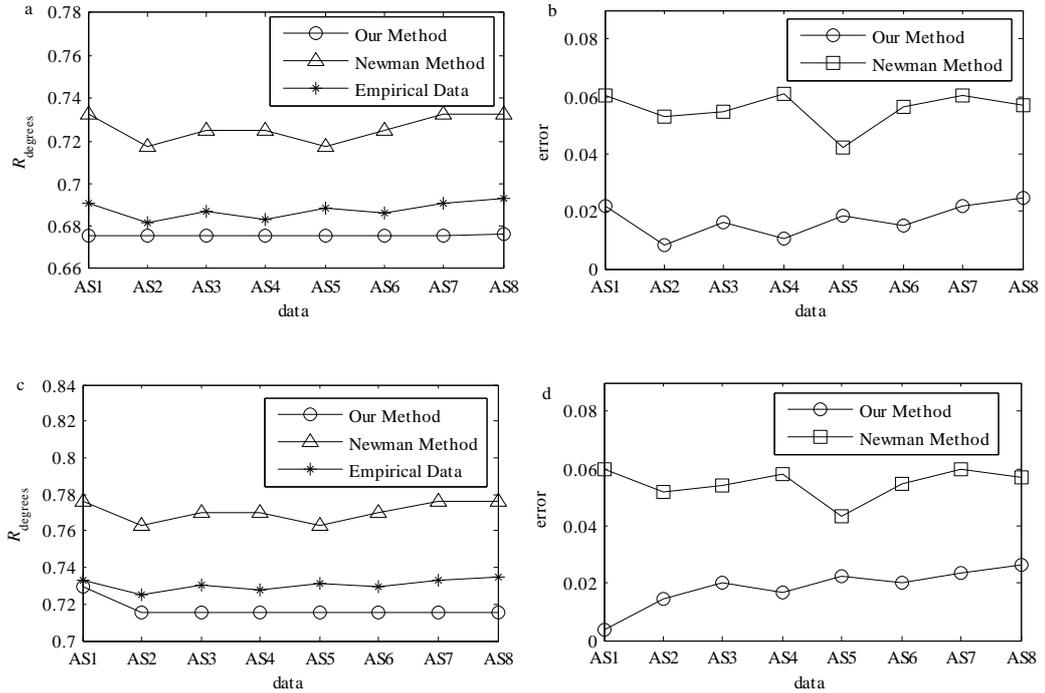

**Fig. 8.** The fraction $R_{degrees}$ of the degrees in the hands of the richest nodes and the errors of the different methods: (a) the values of the fraction $R_{degrees}$ in the hands of top 20% best-connected nodes by the different methods, (b) the values of the errors of the different methods if $R_{nodes}$=20%, (c) the values of the fraction $R_{degrees}$ in the hands of the richest 27% nodes by the different methods, (d) the values of the errors of the different methods if $R_{nodes}$=27%.

# 7. Conclusions

We studied firstly the mathematic model of Internet's scale-free topology. Secondly, we discussed the average degree $<k>$, and the ratios $r_{kmin}$ and $r_{kmax}$ of the $k_{min}$-degree nodes and the $k_{max}$-degree nodes. Thirdly, we discussed the fractions $R_{degrees}$ of the degrees in the hands of the top $R_{nodes}$ best-connected (richest) nodes based on our model. We compared the theoretical and empirical results based on the eight sets of the empirical data extracted from the data source of the AS-level Internet topology: BGP.

We further obtained the calculation formulas of the average degree $<k>$, the ratios of $r_{kmin}$, $r_{kmax}$ of Internet's scale-free topology. We observed many traightforward dependencies: the value of $<k>$ is larger for smaller exponent $\lambda$ and larger values of the minimum degree $k_{min}$, and of the maximum degree $k_{max}$. The value of $r_{kmin}$ is larger for larger $\lambda$ value and smaller $k_{min}$ value and $k_{max}$ value. The value of $r_{kmax}$ is larger for smaller $\lambda$ value and $k_{max}$ value and larger $k_{min}$ value. In addition, we found three new power laws: the increase rate of $<k>$ has the power-law decay with the increase of the $k_{min}$ value. The value of $r_{kmin}$ has the power-law decay with the increase of the $k_{min}$ value. The value of $r_{kmax}$ has the power-law decay with the increase of the $k_{max}$ value. We presented a function relation of $R_{degrees}$ and $R_{nodes}$ and proved theoretically the phenomenon of most of the wealth (degrees or links) in the hands of the few richest people (hubs). We further revealed the '73/27 rule' of Internet's scale-free topology.

We calculated and analysed the theoretical and empirical values of $<k>$, $r_{kmin}$, $r_{kmax}$ and $R_{degrees}$ (if $R_{nodes}$=20% or 27%) of the eight sets of the empirical data (BGP) of the AS-level Internet topology. The results show that our methods are rigorous and effective.

Our results and analysis will improve the scalable routing, search and navigation in the scale-free AS-level Internet, and be fit for other complex networks, such as router-level Internet, WWW, E-mail networks, P2P networks, cellphone networks, neural networks, and so on.

# Acknowledgments

We thank Dr. S Zhou of London University and Professor W. Xiao of South China University of Technology for research support and fruitful discussions. We thank the anonymous reviewers for their insightful comments that have improved the paper considerably.

# References

[1]  Dorogovtsev S N and Goltsev A V 2008 *Rev. Mod. Phys.* **80** 1275

[2]  Castellano C, Fortunato S and Loreton V 2009 *Rev. Mod. Phys.* **81** 591

[3]  Faloutsos M, Faloutsos P and Faloutsos C 1999 *SIGCOMM Comput. Commun. Rev.* **29** 251

[4]  Jaiswal S, Rosenberg A L and Towsley D 2004 *Proceedings of the 12th IEEE International Conference on Network Protocols*, Amherst, USA , October 5-8, 2004 p294

[5]  Zhou S and Mondragon R J 2004 *IEEE Commun. Lett.* **8** 180

[6]  Newman M E J 2002 *Phys. Rev. Lett.* **89** 208701

[7]  Song C, Havlin S and Makse H A 2005 *Nature* **433** 392

[8]  Milo R, Shen-Orr S, Itzkovitz S, Kashtan N, Chklovskii D and Alon U 2002 *Science* **298** 824

[9]  Watts D J and Strogatz S H 1998 *Nature* **393** 440

[10] Barabási A L and Albert R 1999 *Science* **286** 509

[11] Li X and Chen G R 2003 *Physica A* **328** 274

[12] Ravasz E, Somera A L, Mongru D A, Oltvai Z N and Barabási A-L 2002 *Science* **297** 1551

[13] Liu H, Lu J -A, Lü J and Hill D J 2009 *Automatica* **45** 1799

[14] Lü J and Chen G 2005 *IEEE T. Automat. Contr.* **50** 841

[15] Zhou J, Lu J -A and Lü J 2006 *IEEE T. Automat. Contr.* **51** 652

[16] Lü J, Yu X and Chen G 2004 *Physica A* **334** 281

[17] Li X, Wang X F and Chen G 2004 *IEEE T. Circuits-I* **51** 2074

[18] Li Y and Liu Z R 2010 *Chin. Phys. B* **19** 110501

[19] Pastor-Satorras R and Vespignani A 2001 *Phys. Rev. Lett.* **86** 3200

[20] Moreno Y, Gómez J B and Pacheco A F 2002 *Europhys. Lett.* **58** 630

[21] Watts D J, Dodds P S and Newman M E J 2002 *Science* **296** 1302

[22] Adamic L A, Lukose R M, Puniyani A R, Huberman B A 2001 *Phys. Rev. E* **64** 046135

[23] Krioukov D, Papadopoulos F, Boguna M and Vahdat A 2009 *ACM SIGMETRICS Performance Evaluation Review*, **37** 15

[24] Zhang G Q, Zhang G Q, Yang Q F, Cheng S Q and Zhou T 2008 *New J. Phys.* **10** 123027

[25] Cohen R and Havlin S 2003 *Phys. Rev. Lett.* **90** 058701

[26] Mahadevan P, Krioukov D, Fomenkov M, Huffaker B, Dimitropoulos X, Claffy K and Vahdat A 2006 *SIGCOMM Comput. Commun. Rev.* **36** 17

[27] Albert R and Barabási A -L 2000 *Phys. Rev. Lett.* **85** 5234

[28] Bu T and Towsley D 2002 *Proceedings of INFOCOM* New York, USA, June 23-27, 2002 p638

[29] Zhou S and Mondragón R J 2004 *Phys. Rev. E* **70** 066108

[30] Park S, Pennock D M and Giles C L 2004 *Proceedings of INFOCOM* Hong Kong, China, March 7-11, 2004 p1616

[31] Bar S, Gonen M and Wool A 2004 *LNCS* **3015** 53

[32] Chen G, Fan Z P and Li X 2005 *Complex Dynamics in Communication Networks* (Berlin: Springer-Verlag) p174

[33] Winick J and Jamin S http://topology.eecs.umich.edu/inet/ [2010-08-10]

[34] Medina A, Lakhina A, Matta I and Byers J 2001 *Proceedings of the MASCOTS* Cincinnati, USA, August 15-18, 2001 p346

[35] Porekar J http://www-f1.ijs.si/~rudi/sola/Random_Networks.pdf [2010-8-14]

[36] Wang J W and Rong L L 2009 *Phys. A* **388** 1289

[37] Pastor-Satorras R and Vespignani A 2002 *Phys. Rev. E* **65** 035108R

[38] Santiago A and Benito R M 2008 *Phys. A* **387** 2365

[39] Newman M E J 2005 *Contemp. Phys.* **46** 323

[40] Cohen R, Erez K, Ben-Avraham D and Havlin S 2000 *Phys. Rev. Lett.* **85** 4626

[41] Clauset A, Rohilla Shalizi C and Newman M E J 2009 *SIAM Review* **51** 661

[42] Goldstein M L, Morris S A and Yena G G 2004 *Eur. Phys. J. B* **41** 255

[43] Xu H L, Deng X H and Zhang L M 2010 *Computer engineering and applications* **46** 77 (in Chinese)

[44] Xiao W J, Peng L M and Parhami B 2009 *LNICST* **4** 118

[45] Arenas A, Cabrales A, Diaz-Guilera A, Guimera R and Vega-Redondo F 2002 *Proceedings of 18th Sitges Conference on Statistical Mechanics: Statistical Mechanics of Complex Networks*, Sitges, Spain, June 10-14, 2002 p175

[46] Zhang G Q, Wang D and Li G J 2007 *Phys. Rev. E* **76** 017101

[47] Boguñá M, Krioukov D and Claffy K 2009 *Nat. phys.* **5** 74

[48] Zhang L M 2010 *Computer engineering and applications* **46** 4 (in Chinese)

[49] Tsaparas, P http://www.cs.helsinki.fi/u/tsaparas/MACN2006/data-code.html [2010-8-14]